\documentclass[preprint]{aastex631}

\submitjournal{ApJL}

\usepackage[nolist,nohyperlinks]{acronym}
\usepackage[acronym]{glossaries}

\acrodef{gce}[GCE]{galactic chemical evolution}
\acrodef{ccsn}[CCSN]{core collapse supernova}
\acrodefplural{ccsn}[CCSNe]{core collapse supernovae}
\acrodef{mc}[MC]{Monte Carlo}
\acrodef{agb}[AGB]{asymptotic giant branch}
\acrodef{ism}[ISM]{interstellar medium}
\acrodef{sproc}[\textit{s}-process]{slow neutron capture process}

\usepackage{wasysym}
\usepackage{tikz}
\usetikzlibrary{shapes.geometric, arrows, positioning}

\newcommand{\iso}[2]{$^{#1}${#2}}
\def \msun {M_{\odot}}
\newcommand{\del}[3]{\delta ^{#1}\text{#2}_{#3}}
\newcommand{\dsi}[1]{$\del{#1}{Si}{28}$}
\newcommand{\uat}[2]{\href{http://astrothesaurus.org/uat/#2}{#1 (#2)}}

\begin{document}

\title{Silicon Isotopic Composition of Mainstream Presolar SiC Grains Revisited: The Impact of Nuclear Reaction Rate Uncertainties}

\correspondingauthor{Reto Trappitsch}
\email{reto.trappitsch@epfl.ch}

\author[0009-0008-6833-7300]{Hung Kwan Fok}
\altaffiliation{Present address: Morton K. Blaustein Department of Earth \& Planetary Sciences, Johns Hopkins University, Olin Hall, 3300 San Martin Drive, Baltimore, MD 21218}
\affiliation{Department of Physics\\
Brandeis University\\
Abelson-Bass-Yalem 107\\
Waltham, MA 02453, USA}
\affiliation{NuGrid collaboration, \url{https://nugridstars.org}}

\author[0000-0002-9048-6010]{Marco Pignatari}
\affiliation{Konkoly Observatory\\
Research Centre for Astronomy and Earth Sciences\\
HUN-REN\\
Konkoly Thege M. út 15-17\\
Budapest 1121, Hungary}
\affiliation{CSFK\\
MTA Centre of Excellence\\
Konkoly Thege Miklós út 15-17\\
Budapest 1121, Hungary
}
\affiliation{E. A. Milne Centre for Astrophysics\\
University of Hull\\
Cottingham Road\\
Kingston upon Hull, HU6 7RX, UK
}
\affiliation{NuGrid collaboration, \url{https://nugridstars.org}}

\author[0000-0002-9986-8816]{Benoît Côté}
\affiliation{Konkoly Observatory\\
Research Centre for Astronomy and Earth Sciences\\
HUN-REN\\
Konkoly Thege M. út 15-17\\
Budapest 1121, Hungary}
\affiliation{NuGrid collaboration, \url{https://nugridstars.org}}

\author[0000-0002-0924-236X]{Reto Trappitsch}
\affiliation{Laboratory for Biological Geochemistry\\
School of Architecture, Civil \& Environmental Engineering\\
École Polytechnique Fédérale de Lausanne\\
GR C2 505, Station 2\\
1015 Lausanne, Switzerland}
\affiliation{Department of Physics\\
Brandeis University\\
Abelson-Bass-Yalem 107\\
Waltham, MA 02453, USA}
\affiliation{NuGrid collaboration, \url{https://nugridstars.org}}

\begin{abstract}
Presolar grains are stardust particles that condensed in the ejecta or in the outflows of dying stars and can today be extracted from meteorites. 
They recorded the nucleosynthetic fingerprint of their parent stars and thus serve as valuable probes of these astrophysical sites. 
The most common types of presolar silicon carbide grains (called mainstream SiC grains) condensed in the outflows of asymptotic giant branch stars. 
Their measured silicon isotopic abundances are not significantly influenced by nucleosynthesis within the parent star, but rather represents the pristine stellar composition. 
Silicon isotopes can thus be used as a proxy for galactic chemical evolution. 
However, the measured correlation of \iso{29}Si/\iso{28}{Si} versus \iso{30}{Si}/\iso{28}{Si} does not agree with any current chemical evolution model. 
Here, we use a Monte Carlo model to vary nuclear reaction rates within their theoretical or experimental uncertainties and process them through stellar nucleosynthesis and galactic chemical evolution models to study the variation of silicon isotope abundances based on these nuclear reaction rate uncertainties. 
We find that these uncertainties can indeed be responsible for the discrepancy between measurements and models and that the slope of the silicon isotope correlation line measured in mainstream SiC grains agrees with chemical evolution models within the nuclear reaction rate uncertainties.  
Our result highlights the importance of future precision reaction rate measurements for resolving the apparent data-model discrepancy. 

\end{abstract}

\keywords{
\uat{Circumstellar grains}{239}
\uat{Stellar nucleosynthesis}{1616}
\uat{Galaxy chemical evolution}{580};
\uat{Reaction rates}{353};
\uat{Isotopic abundances}{867}
}

\section{Introduction} \label{sec:intro}
Presolar grains are small particles that condensed in the outflows of dying stars. 
After they drifted through the \ac{ism}, some of them were incorporated into the first condensing solids in the solar nebula and thus into meteorite precursors. 
Today, these grains can be extracted from primitive meteorites, i.e., meteorites that were never altered during the 4.567\,Gyr since the solar system formed \citep[e.g.,][]{nittlerAstrophysicsExtraterrestrialMaterials2016}.
Presolar grains recorded the nucleosynthesis fingerprint of their parent stars at the time and in the region where the particles condensed. Analyzing them thus allows us to directly probe the isotopic composition of these astrophysical sites.

Currently, the best-studied presolar phase are SiC grains, mainly since they occur in sizes of up to several micrometers and are thus relatively large compared to, e.g., presolar silicates, and since their chemical makeup is strong enough such that they can be separated from the meteoritic host material by etching with strong acids \citep{amariInterstellarGrainsMeteorites1994}. 
This latter point is also highly effective in removing any solar system contamination from these particles, such that the pristine stellar material can be measured without significant dilution with solar system material \citep{liuBariumIsotopicComposition2014}. The majority of presolar SiC grains, known as mainstream grains, condensed in the outflows of low-mass stars (up to a few solar masses) when they underwent the \acf{agb} phase at the end of their lives. These stars have been observed to produce \ac{sproc} nuclei \citep[e.g.,][]{merrillSpectroscopicObservationsStars1952} and the same observations have also been confirmed in SiC mainstream grains \citep{savinaExtinctTechnetiumSilicon2004}, showing the provenance of these grains. The \ac{sproc} mainly produces isotopes heavier than iron \citep[e.g.,][]{gallinoEvolutionNucleosynthesisLowmass1998}. 
Below iron, presolar SiC mainstream grains show isotopic compositions of certain elements, e.g., silicon and titanium, that represent the original composition the parent star started out with.
Indeed, the intrinsic nucleosynthesis processes in the grain's parent \ac{agb} stars are not expected to introduce significant anomalies in the isotopic composition of these species \citep{Clayton&Timmes_1997,Alexander_1999}.

The silicon isotopic ratios of the majority of the SiC mainstream grains are enriched in the secondary isotopes \iso{29}{Si} and \iso{30}{Si} compared to the solar system. Within a classic \ac{gce} scenario \citep[e.g.,][]{kobayashiOriginElementsCarbon2020}, the silicon isotopes of these grains would indicate that their parent stars were more evolved in terms of \ac{gce} compared to the solar system. Furthermore, the correlation of \dsi{29} versus \dsi{30}\footnote{
Throughout this manuscript, we express silicon isotope ratios as $\delta$-values. These are defined as
\begin{equation}
    \delta ^{i}\mathrm{Si}_{28} 
        = \delta \left( \frac{^{i}\mathrm{Si}}{^{28}\mathrm{Si}}\right) 
        = \left[ \frac{ \left( ^{i}\mathrm{Si}/^{28}\mathrm{Si} \right)_\text{m}}{\left(^{i}\mathrm{Si}/^{28}\mathrm{Si}\right)_\odot} 
            - 1\right] \cdot 1000 \qquad (\permil).
        \label{eqn:delta_notation}
\end{equation}
Here, $i$ is the mass number of the minor silicon isotopes ($i \in \{29,30\}$). Subscript ``m'' and ``$\odot$'' refer to the model output/presolar grain measurements and solar abundances of the given isotopes, respectively. Overall, $\delta$-values represent the deviation of a modeled or measured value from the solar isotope ratio \citep{loddersRelativeAtomicSolar2021} in permil.
}
as measured in presolar grains is significantly larger than predictions of \ac{gce} models. While the linear correlation along the mainstream line predicts a slope of $1.342 \pm 0.004$ \citep{stephan2024pgd}, typical homogeneous \ac{gce} models predict correlation lines with a slope of around unity \citep[e.g.,][]{Clayton&Timmes_1997}. For clarity, we will refer subsequently to the SiC mainstream grain correlation line as the ``mainstream line'' and the analogous \ac{gce} model predicted correlation line as the ``\ac{gce} line''. Numerous studies have sought to resolve this model-data discrepancy. \citet{Clayton&Timmes_1997} proposed an explanation for the slope of the mainstream line, suggesting that the Sun is significantly enriched in \iso{30}{Si} compared to the typical ISM. However, this explanation requires the parent stars of the mainstream grains to experience a more extensive dredge-up of He-shell materials, which contradicts the \iso{12}{C}/\iso{13}{C} ratios in the mainstream grains \citep{Alexander_1999}. \citet{Lugaro1999} illustrated that a heterogeneous \ac{gce} predicts a slope higher than the typical homogeneous \ac{gce} model but still lower than the slope of the mainstream line. An alternative theory was introduced by \citet{clayton2003}, who postulated that the silicon isotopic ratio in the mainstream grains might represent the mixing line of a merger between a dwarf galaxy and the Milky Way disk before the formation of the Solar System. However, this model remains speculative and lacks concrete evidence.

In this study, we approach the problem from a nuclear physics perspective. Previous studies suggested that uncertainties in nuclear reaction rates might explain the discrepancy between the \ac{gce} model predictions and mainstream grain measurements \citep{Timmes_1996}. \citet{Timmes_1996} came to this conclusion by analyzing GCE models in comparison with presolar SiC mainstream grains. They could not find an agreement between model and data but speculated that nuclear reaction rate uncertainties could play a major role.

While silicon is mainly produced in \acp{ccsn} \citep[e.g.,][]{kobayashiOriginElementsCarbon2020}, presolar grains from \ac{agb} stars, e.g., mainstream SiC grains, trace the \ac{gce}-averaged signal of all \acp{ccsn}. 
\citet{zinner:06} studied the silicon isotopic composition of SiC Z grains, which were hypothesized to have formed in low-mass, low-metallicity \ac{agb} stars \citep{hoppe:1997}. 
These grains are characterized by elevated \iso{30}{Si}/\iso{28}{Si} ratios compared to SiC mainstream grains. 
\citet{zinner:06} concluded for the used \ac{agb} star nucleosynthesis models 
that using the neutron-capture reaction rates by \citet{guber} yields a better agreement with the SiC Z grain measurements than when using the rates by \citet{bao}.

\citet{Hoppe_2009} analyzed presolar SiC X grains, which condensed in the ejecta of \acp{ccsn}. 
Each of these samples thus only probes one individual \ac{ccsn} event and the grain's silicon isotopic composition does not represent the average \ac{gce} signal. By analyzing SiC X grains, \citet{Hoppe_2009} found that they can explain all isotopic measurements except for the abundance of \iso{29}{Si} rather well with existing \ac{ccsn} models. However, in order to match the \iso{29}{Si} measurements they needed to boost its production in the model. Enhancing the \iso{26}{Mg}$(\alpha,n)$\iso{29}{Si} reaction rate in the explosive carbon and oxygen burning calculations by a factor of three led to a good match of measurements and models. 

The studies by \citet{zinner:06} and \citet{Hoppe_2009} highlight the high potential 
impact of nuclear uncertainties on nucleosynthesis models, and consequently, we expect a similar effect on \ac{gce} models of silicon isotopes. Notably, the increased \iso{29}{Si}/\iso{30}{Si} ratio shown in \citet{Hoppe_2009} might elevate the \ac{gce} line and potentially resolve the model-data discrepancy. In this paper, we quantify, for the first time, the impact of nuclear uncertainties on the \ac{gce} of silicon isotopes using \iac{mc} framework.

\section{Methods and model} \label{sec:model}

In this section we first discuss the uncertainties of the 
relevant nuclear reaction rates leading to the production and destruction of silicon isotopes. Then, we present models used to estimate the impact of nuclear reaction rate uncertainties on the \ac{gce} of silicon isotopes.

\subsection{Nuclear reaction rates and their uncertainties}

Nuclear reactions rates are a key input for %
nucleosynthesis %
in stellar evolution models. 
Their uncertainties %
are propagating through the stellar %
simulations and into the results of \ac{gce} %
(Figure~\ref{fig:pipeline}). 
In this work we focus
solely on the effects of nuclear uncertainties on the \ac{gce} of silicon isotopes. 

\begin{table}[bt]
    \centering
    \caption{{Uncertainty factors used in this work for considered nuclear reaction rates. For the upper and lower limits of the considered uncertainties, the reaction rates are multiplied by $U^{\rm hi}$ and $U^{\rm lo}$, respectively. For experimentally determined reaction rates, $2\sigma$ uncertainties were used as the upper and lower limits. For theoretical rates, we use the same uncertainties as \citet{Rauscher_2016}.}}
\begin{tabular}{c c c c}
        \hline
          Reaction &    $U^{\rm hi} $ & $U^{\rm lo}$ & Type\footnote{exp: experimentally determined rates; th: theoretically determined rates}\\
        \hline
          \iso{28}{Si}$(n, \gamma)$\iso{29}{Si} & 1.18\footnote{Uncertainties as reported by \citet{guber}\label{fn:guber}} / 1.21\footnote{Uncertainties as reported by \citet{bao}\label{fn:bao}} & 0.82\footref{fn:guber} / 0.79\footref{fn:bao} & exp \\
        \iso{29}{Si}$(n, \gamma)$\iso{30}{Si} & 1.20\footref{fn:guber} / 1.23\footref{fn:bao} & 0.80\footref{fn:guber} / 0.77\footref{fn:bao} & exp \\
        \iso{30}{Si}$(n, \gamma)$\iso{31}{Si} & 1.36\footref{fn:guber} / 1.18\footref{fn:bao} & 0.64\footref{fn:guber} / 0.82\footref{fn:bao} & exp \\
        \iso{24}{Mg}$(\alpha, \gamma)$\iso{28}{Si} & 1.3  & 0.7 & exp \\
        \iso{25}{Mg}$(\alpha, n)$\iso{28}{Si} & 2.000  & 0.100 & th \\
        \iso{26}{Mg}$(\alpha, n)$\iso{29}{Si} & 2.000  & 0.100 & th \\
        \iso{32}{S}$(n, \alpha)$\iso{28}{Si} & 2.000  & 0.100 & th \\
        \iso{33}{S}$(n, \alpha)$\iso{29}{Si} & 2.000  & 0.100 & th \\
        \hline
    \end{tabular}
    \label{tab:rate-variation}
\end{table}

We concentrate our uncertainty study on the specific region in \acp{ccsn} where silicon isotopes are produced. Mass coordinates for the individual zones are given in Appendix~\ref{app:o-burning}. 
The main paths to produce and destroy silicon in this region are via the reactions shown in Table~\ref{tab:rate-variation}. To determine these main reactions, we consider the contribution of each individual reaction rate to the sum total of reactions in the analyzed stellar zones. 
The reactions {named} in Table~\ref{tab:rate-variation} are the ones that change the silicon isotopic composition by $\delta_i$ such that $|\delta_i/\delta_{max}|\geq 0.001$ where $\delta_{max} = \max(\{\delta_i: i=\text{a reaction in the network}\})$. 
 {For each nuclear reaction rate, the table presents the upper and lower limit of the associated uncertainties.} For measured reaction rates we use $\pm 2\sigma$ for the uncertainty in our model, where $\sigma$ represents the analytical one standard deviation uncertainty. For reaction rates that were not measured and for which thus only theoretical values and no uncertainties exist, we employ the same uncertainties that were used by \citet{Rauscher_2016}. 

 For the \iso{28,29,30}Si($n,\gamma$) {nuclear reaction rates we performed nucleosynthesis calculations using two distinct sets of rates, namely} the ones from \citet{guber} and from \citet{bao}. While the \citet{guber} rates were used in the nucleosynthesis calculations by,
 e.g., \citet{zinner:06} and \citet{Ritter_2018}, these rates are still controversial in particular for \iso{30}Si. 
 We therefore use the rates by \citet{guber} as the standard case to discuss the effect of nuclear reaction rate uncertainties on the \ac{gce} of silicon isotopes and compare our final results with a second set of simulations using the rates by \citet{bao}.
 Uncertainties for both sets are given in Table \ref{tab:rate-variation}.
 
\subsection{Models}

\begin{figure}[bt]
    \centering
     \includegraphics[scale=0.75]{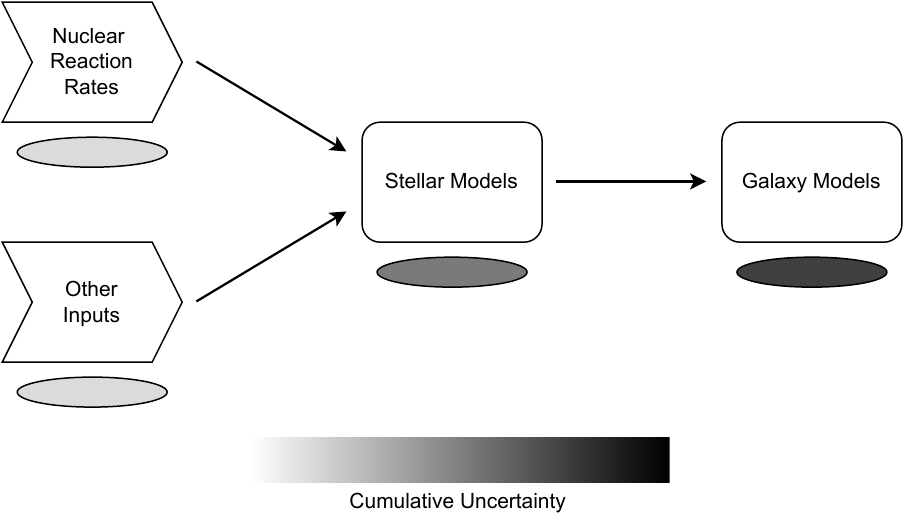}
    \caption{Schematic overview of the \ac{gce} pipeline that we employ in this work to study the effects of nuclear reaction rate uncertainties on the evolution of silicon isotopes in the Milky Way. The gray scale shows schematically how uncertainties propagate through the model and how small initial uncertainties can have a large effect on predicting isotopic compositions.}
    \label{fig:pipeline}
\end{figure}

We developed a nucleosynthesis-\ac{gce} \ac{mc} framework to quantify the effect of nuclear uncertainties on the stellar yields of their respective stars and consequently on the results of \ac{gce}. Figure~\ref{fig:pipeline} shows an overview of the parts that are important for this work. 
The framework can be separated into two sub-models: the stellar model and the galaxy model. 

The stellar evolution calculation is the starting point of our framework. Here, we utilize the NuGrid stellar models by \cite{Ritter_2018}, whose 1D stellar evolution calculations were done with the \texttt{MESA} stellar evolution code \citep[rev. 3372;][]{Paxton2011,Paxton2013,Paxton2015}. \Acl{ccsn} explosions are based on the semi-analytical approach presented in \citet{Pignatari_2016}. The thermodynamic and structural information of the stellar evolution calculation and the \ac{ccsn} explosion model are then extracted and post-processed using the multi-zone nuclear reaction network \texttt{mppnp} for the nucleosynthesis calculation using a full nuclear reaction rate network \citep{Ritter_2018}.  

The resulting stellar yields are the input for the galaxy model. Here, we use the 2-zone semi-analytical galaxy model \texttt{OMEGA+} {\citep{cote17_omega,cote18_omega+}}. \texttt{OMEGA+} consists of a star-forming region, representing the galaxy, surrounded by a hot gas reservoir filling the dark matter halo of the host galaxy and takes a set of galaxy parameters and stellar yields to calculate the evolution of elemental and isotopic abundances in the galaxy as a function of time. The input parameters of the \texttt{OMEGA+} calculations were calibrated to reproduce the current bulk properties of the Milky Way \citep{omega+setup}. The parameter values are equivalent to the values in the "best" model of \citet{omega+setup} and are identical to the values used in by \citet{womack2023}.
{Different sets of CCSN yields are adopted to generate the GCE models presented in this work: namely \citet{Ritter_2018}, \citet{nomoto13} and \citet{lc18}. The largest stellar masses available are M = 25 M$_{\odot}$, 40 M$_{\odot}$ and 120 M$_{\odot}$, respectively. For consistency, the maximum contributing stellar mass adopted for all GCE simulations is M = 120 M$_{\odot}$, which is the largest stellar model provided by \cite{lc18}. For the other GCE models using yields by \cite{Ritter_2018} and \cite{nomoto13}, we use the standard extrapolation setup in \texttt{OMEGA+}, i.e., applying the yields of the most massive model available to all more massive stars.}

{The theoretical and experimental nuclear reaction rates were randomly varied within their uncertainties using a uniform distribution. This approach is identical to the one chosen by \citet{Rauscher_2016}. While experimental reaction rate uncertainties are generally better described using a Gaussian distribution, our results show that systematic differences between different studies play a larger effect than reaction rate uncertainties}. Nucleosynthesis within silicon production \ac{ccsn} zone for each star is then calculated. This allows us to propagate the uncertainties of the nuclear reaction rates to the the total stellar yields. We then input the modified stellar yield into the galaxy model to determine the uncertainties of the GCE model. 
This means that for each \ac{mc} iteration, a set of rate variation factors $\mu_{\text{rate}}$ for the rates of all the reactions shown in Table \ref{tab:rate-variation} are randomly generated according to a uniform distribution within the uncertainty limits defined in Table \ref{tab:rate-variation}. In order to perform a statistically sufficient amount of \ac{mc} calculations, post-processing the full stellar nucleosynthesis for each set of modified reaction rates is computationally prohibitive. %
Therefore, for all stellar models we take representative 1D temperature, plasma-density trajectories from their respective explosive silicon production zones, ranging from mass-coordinates $m_a$ to $m_b$ (details are given in Appendix~\ref{app:trajectories}). The subscript represents the index of the computational grid of the \texttt{MESA} stellar evolution code, where 0 represents the center of the star. Each trajectory traces temperature $T$ and plasma density $\rho$ at mass-coordinate $m_n$ over time from the onset of core-collapse ($t=0$) to explosion completion ($t=t_f$). 
The isotope yield scale factor, i.e., the enhancement factor with respect to the trajectory with unmodified reaction rates, $\lambda_{n}$ at each trajectory's mass coordinate $m_{n}$ is then computed. For any $m_i$, the scale factor $\lambda_i$ is furthermore determined by linear interpolation between the closest neighbors.
For $i<a$ or $i>b$, $\lambda_i=1$. The integrated stellar nucleosynthesis yield for an isotope is then computed by summation as follows:
\begin{align}
m = \sum_{i=k}^{M-1} \lambda_i X_i \Delta m_i
\end{align}
Here, $m_k$ is the mass cut derived from \cite{Fryer_2012}, $X_i$ is the isotopic mass fraction in the stellar nucleosynthesis models in \cite{Ritter_2018}, and $\Delta m_i = m_{i+1}-m_i$. The yield variation factor $\nu_{\text{yield}}$ is then calculated as the mass of the total modified yield divided by the mass of the unmodified yield \citep{Ritter_2018}.
These rate variation factors are mapped to a set of yield variation factors, $\nu_{\text{yield}}$, for all massive star models and all silicon isotopes \iso{28}{Si}, \iso{29}{Si}, and \iso{30}{Si}.
Finally, the set of yield variation factors $\nu_{\text{yield}}$ is passed into the \texttt{OMEGA+} to calculate the \ac{gce} of silicon isotopes corresponding to the set of rate variation factors $\mu_{\text{rate}}$.  This whole process was then repeated for the next set of randomly varied nuclear reaction rates.

{Overall, we propagated nuclear reaction rate uncertainties via the described \ac{mc} approach through eight stellar models. Specifically, we modeled four masses (12, 15, 20, and $25\,M_\odot$) at two different metallicities (Z=0.01 and 0.02). The models by \citet{Ritter_2018} are based on initial abundances scaled to the solar abundances by \citet{gn93} (Z=0.019), and the $Z=0.02$ cases are the closest models to this solar metallicity. More modern solar references are anyway included between Z=0.01 and Z=0.02 \citep[e.g.,][]{loddersRelativeAtomicSolar2021, magg2022}. The models at these metallicities are the best to use to explore the impact of nuclear uncertainties on the production in CCSNe of Si isotopes. For $^{28}$Si, its production is primary and therefore the initial metallicity of the models does not significantly affect the results. But $^{29}$Si and $^{30}$Si are mostly secondary-like products, and their production tend to increase with the initial metallicity of the star \citep[e.g.,][]{Ritter_2018, lc18}. Therefore, for the GCE to produce the Si solar abundances and the population of presolar grains, the mildly sub-solar and solar-like metallicity models will be representative of all the Si isotopes.  
For the \ac{gce} calculations, the full suite of stellar models from \citet{Ritter_2018} are considered. Addtitional masses required for the \ac{gce} calculation are inter-/extrapolated from nearby models as a function of metallicty \citep[see][]{cote17_omega}.}

\section{Result \& Discussion} \label{sec:result}
\subsection{Stellar Nucleosynthesis Uncertainties}\label{sec:stellar-uncertainties}
To determine the impact of nuclear reaction rate uncertainties on the \ac{gce} of silicon isotopes, we performed 10000 \ac{mc} simulations to vary the nuclear reaction rates that lead to the production and destruction of silicon isotopes following the procedure described in Section \ref{sec:model}. Convergence of our \ac{mc} model is discussed in Appendix~\ref{app:mc_convergence}. %
\begin{figure}[tb]
    \centering
    \includegraphics[width=0.5\textwidth]{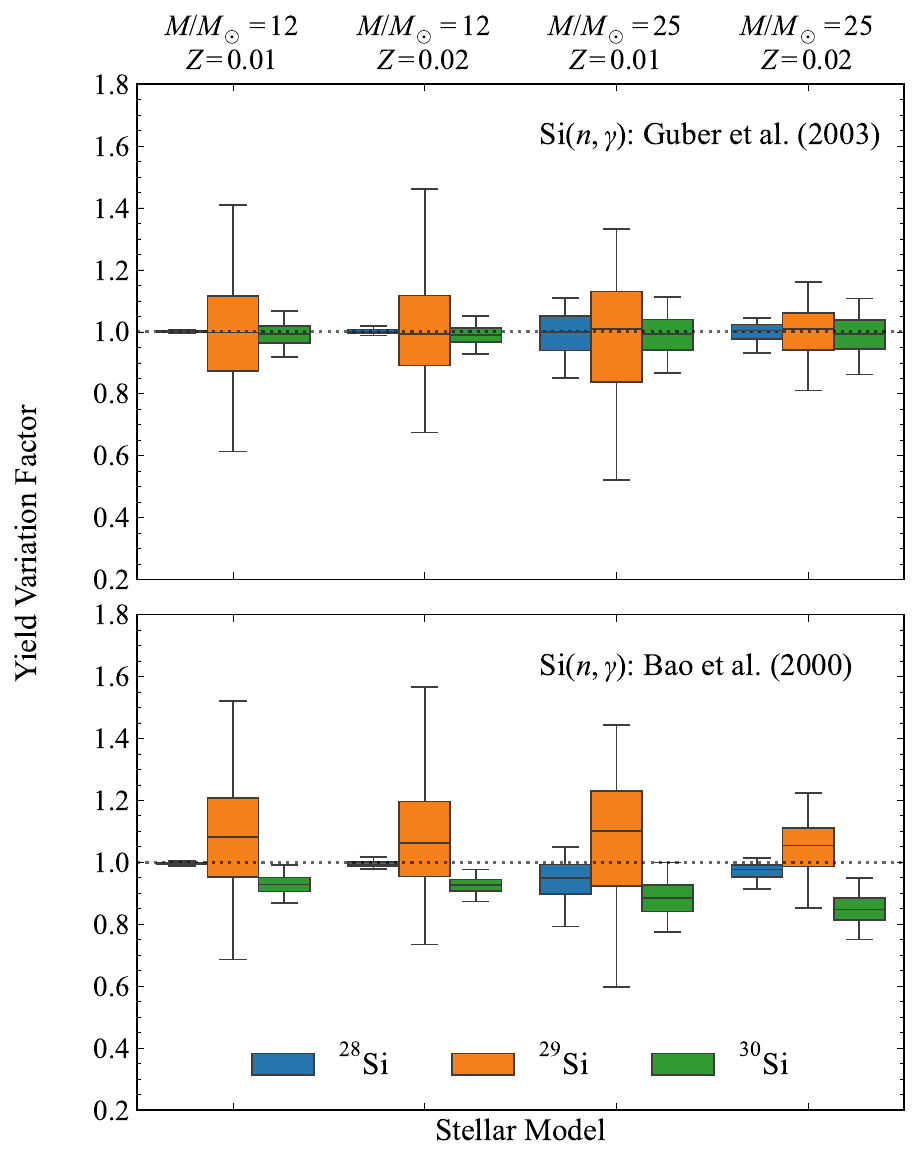}
    \caption{Silicon isotope integrated stellar yield variations for the $12\,M_\odot$ and $25\,M_\odot$ stars at $Z=0.01$ and $Z=0.02$. The dotted horizontal line indicates the unmodified stellar yields. {The variations in the integrated yields are shown as boxes and bars for the 50\% and 95\% confidence interval, respectively. The top panel shows the yield variation factors using the Si$(n,\gamma)$ rates by \citet{guber}, the bottom panel using the rates by \citet{bao}. Note that the dashed line at unity represents in both cases the original stellar models, which used the Si$(n,\gamma)$ rates by \citet{guber}.} Clearly, the largest effect due to nuclear reaction rate uncertainties are seen in \iso{29}{Si}.}
    \label{fig:yield-variation}
\end{figure}
The key inputs for the \ac{gce} model are the overall stellar yields and thus, the uncertainty of these models is determined by the integrated variations in silicon isotopic yields. Figure \ref{fig:yield-variation} shows the variations of the silicon isotopic massive star yields due to nuclear reaction rate uncertainties for selected massive star models {using Si$(n,\gamma)$ rates by \citet{guber} (top) and \citet{bao} (bottom). Details on variations with the stellar zones are provided in Appendix~\ref{app:silicon_production_variation}}. Boxes in Figure~\ref{fig:yield-variation} depict the 50\% confidence intervals of the variation in integrated stellar yields while bars show the 95\% confidence interval. Clearly, the \iso{29}{Si} yields show the largest variations. While for the $12\,M_\odot$ star, the variation factors of \iso{29}{Si} are more than a factor of 1.3 higher than for all other isotopes, the difference is significantly lower for the $25\,M_\odot$ star. This effect is due to the fallback radius in the $25\,M_\odot$ stars being larger than in the $12\,M_\odot$ stars. The $25\,M_\odot$ stars therefore ejects significantly less silicon from the center and thus recycles less material back into the galaxy. Therefore, silicon modified in the explosion contributes a larger part of the overall yield, which results in the larger variation factors for all silicon isotopes. Note that the variation factors shown in Figure~\ref{fig:yield-variation} are not weighted by the overall yields. Thus, the overall contribution of the individual stars to \ac{gce} cannot be derived from this figure.

\newpage
\subsection{GCE models and propagated uncertainties}\label{sec:results_gce_models_prop_unc}
The stellar structure and nucleosynthesis yields in the models by \citet{Ritter_2018} are heavily affected by the convective O-C shell merger during silicon shell burning, which occurs in some of the stellar models. For example, \citet{ritter-letter}
demonstrates that burning of ingested neon during the shell mergers results in a boost of odd-Z elements and, when propagated through a \ac{gce} model, results in large overproduction factors for these elements of up to {a factor of ten}.
{We find that O-C shell mergers greatly affect the production of Si isotopes in massive stars, and also have a clear impact on our GCE results.}

\begin{figure}
    \centering
    \includegraphics[width=0.7\textwidth]{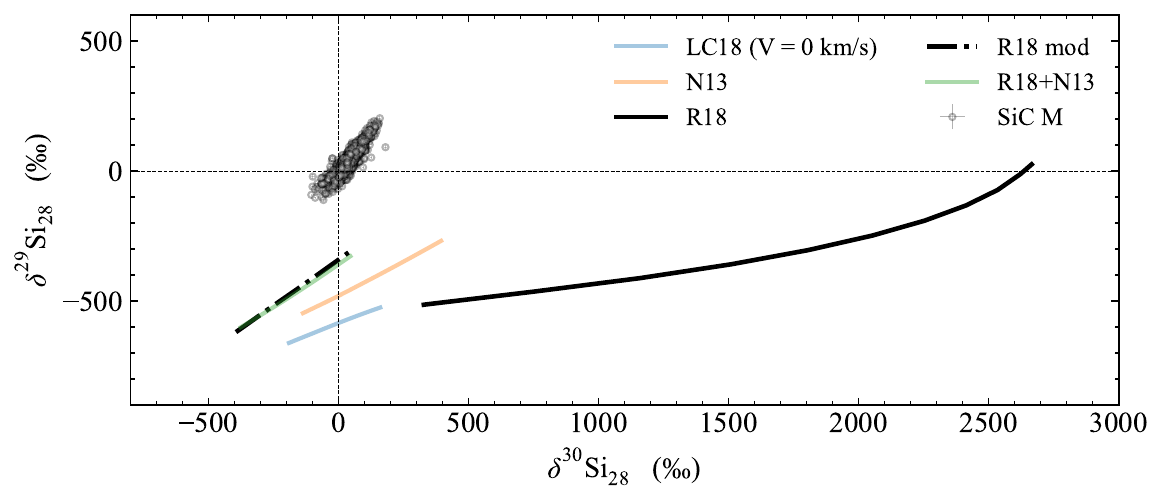}
    \caption{Comparison of \ac{gce} model predictions of silicon isotope evolution using different stellar yield sets with presolar SiC mainstream grains (SiC M). The simulated time range is between 4.2\,Ga prior to solar system formation (left hand side of the model lines) up to solar system formation (right hand side of model lines). Citation key: N13 - \citet{nomoto13}, LC18 - \citet{lc18}, R18 - \citet{Ritter_2018}. Presolar grain measurements are taken from the presolar grain database \citep{stephan2024pgd}.}
    \label{fig:gce-comparison}
\end{figure}

For example, the \iso{30}{Si} yield in the $15\,\msun,\ Z=0.02$ model of \citet{Ritter_2018} is {about ten times} higher than the \iso{30}{Si} yield predicted by other massive star models that do not undergo a convective O-C shell merger \citep{lc18, nomoto13}. %
In Figure~\ref{fig:gce-comparison} we compare the \texttt{OMEGA+} predictions using the stellar yield set of \citet{Ritter_2018} with the predictions using the stellar yield sets of other massive star models by \citet{nomoto13} and \citet{lc18}.
The stellar models of \citet{Ritter_2018} significantly overproduce \iso{30}{Si} compared to other stellar models in \ac{gce}. {However, at the same time they are the sole models that reproduce the measured \iso{29}{Si}/\iso{28}{Si} isotope ratios}. This is a direct result of the O-C shell mergers that occur in the $15\,M_\odot$ and $20\,M_\odot$ stars in these models. On the other hand, \ac{gce} predictions using stellar yields by \citet{nomoto13} and \citet{lc18} generally agree well with each other, {underproducing $^{29}$Si compared to both the Sun and the presolar grains sample. This is a well-known issue found by previous GCE studies focused on the Si isotopes \citep[e.g.,][]{Lugaro1999}.} 

The occurrence of O-C shell mergers and the accuracy of the 1D model predictions are highly uncertain. Other massive star models do not show the occurrence of O-C shell mergers, or they show them at different progenitor masses and/or metallicities. The overall nucleosynthesis impact may also vary strongly between different models \citep[e.g., see recently][]{roberti:24}. %
Moreover, \citet{Ritter_2018} shows that a major O-C shell merger in the $M/\msun =15,\ Z=0.02$ model disappears in higher resolution simulations. %
In general, large-scale asymmetries arising in the 3D calculations of the O-C shell merger in \citet{Andrassy_2019} suggest that 1D spherical symmetric models cannot capture the behavior of the shell merger.

We performed two test cases to demonstrate the effect of the O-C shell merger on silicon isotopes. In the first test case, we substituted the stellar yields of the $M/\msun =15, 20\ , Z=0.01, 0.02$ models, with values given by linear interpolations of the yields of the nearby models 
(i.e., $M/\msun =12, 25\ , Z=0.01, 0.02$). This interpolation is based on the default \texttt{OMEGA+} assumption that yields depend linearly on the initial mass and metallicity of the stellar model. This test case is shown as a dashed-dotted line in Figure~\ref{fig:gce-comparison} labeled ``R18 mod''. In the second test case, we substituted the stellar yields of the $M/\msun =15, 20\ , Z=0.01, 0.02$ models with yields from \citet{nomoto13}, shown as the green line labeled ``R18+N13'' in Figure~\ref{fig:gce-comparison}. We find that the two test cases yield almost identical results with respect to the \ac{gce} of silicon isotopes. Moreover, the resulting \ac{gce} curves of the two test cases show \dsi{30} values that are similar to the \dsi{30} values of the \ac{gce} curves using stellar yield sets by \citet{nomoto13} and \citet{lc18}. 

{Coming back to the comparison with the solar abundances and the Si isotopic ratios in presolar SiC mainstream measurements,} two main issues arise from \ac{gce} simulations:  %
(1) the models do not {reproduce the heavy silicon isotopic compositions measured in presolar SiC mainsteram grains} and (2) the slope of all \ac{gce} models is significantly shallower than the {measured slope}. \citet{Timmes_1996} already discussed these issues and stated that accumulated uncertainties {will always result in models having difficulties to reproduce the precise isotopic composition of presolar grain measurements.} A further issue represents itself in the fact that these grains seem enriched in secondary \iso{29,30}{Si} when compared with solar. Therefore, a simple \ac{gce} model based on the age-metallicity relation will never be able to {explain the measured enhancement in \iso{29,30}{Si} with respect to solar}, since {the model} will, by definition, end up at solar composition at the time of birth of the solar nebula. Only heterogeneous \ac{gce} models \citep[e.g.,][]{Lugaro1999, Nittler_2005} have the potential of predicting the {measured isotope ratios}. \citet{Timmes_1996} however also pointed out that the slope of the predicted \ac{gce} silicon isotope correlation line is approximately one, i.e., significantly lower than the measured mainstream line. They suggested that this discrepancy could likely be due to nuclear reaction rate uncertainties. In fact, we see that the modified \citet{Ritter_2018} models predict a slope of approximately one. Other stellar yields by \citet{nomoto13} and \citet{lc18} result in an even shallower slope, increasing the discrepancy.
 
Using the modified \citet{Ritter_2018} stellar yields (``R18 mod'' in Figure~\ref{fig:gce-comparison}), we calculated the uncertainties of the slope of the silicon \ac{gce} correlation line due to nuclear reaction rate uncertainties. We varied the stellar yields of the $M/\msun =12, 25\ , Z=0.01, 0.02$ models based on the \ac{mc} results shown in Section~\ref{sec:stellar-uncertainties}. {While the individual rates were varied independently from each other, the same rate variations for any given run are consistently used across the different stellar masses and metallicities that we modeled.} The stellar yields of the $M/\msun =15, 20\ , Z=0.01, 0.02$ models were then interpolated based on the modified yields of the $M/\msun =12, 25\ , Z=0.01, 0.02$ models. 
\begin{figure}
    \centering
    \includegraphics[width=\textwidth]{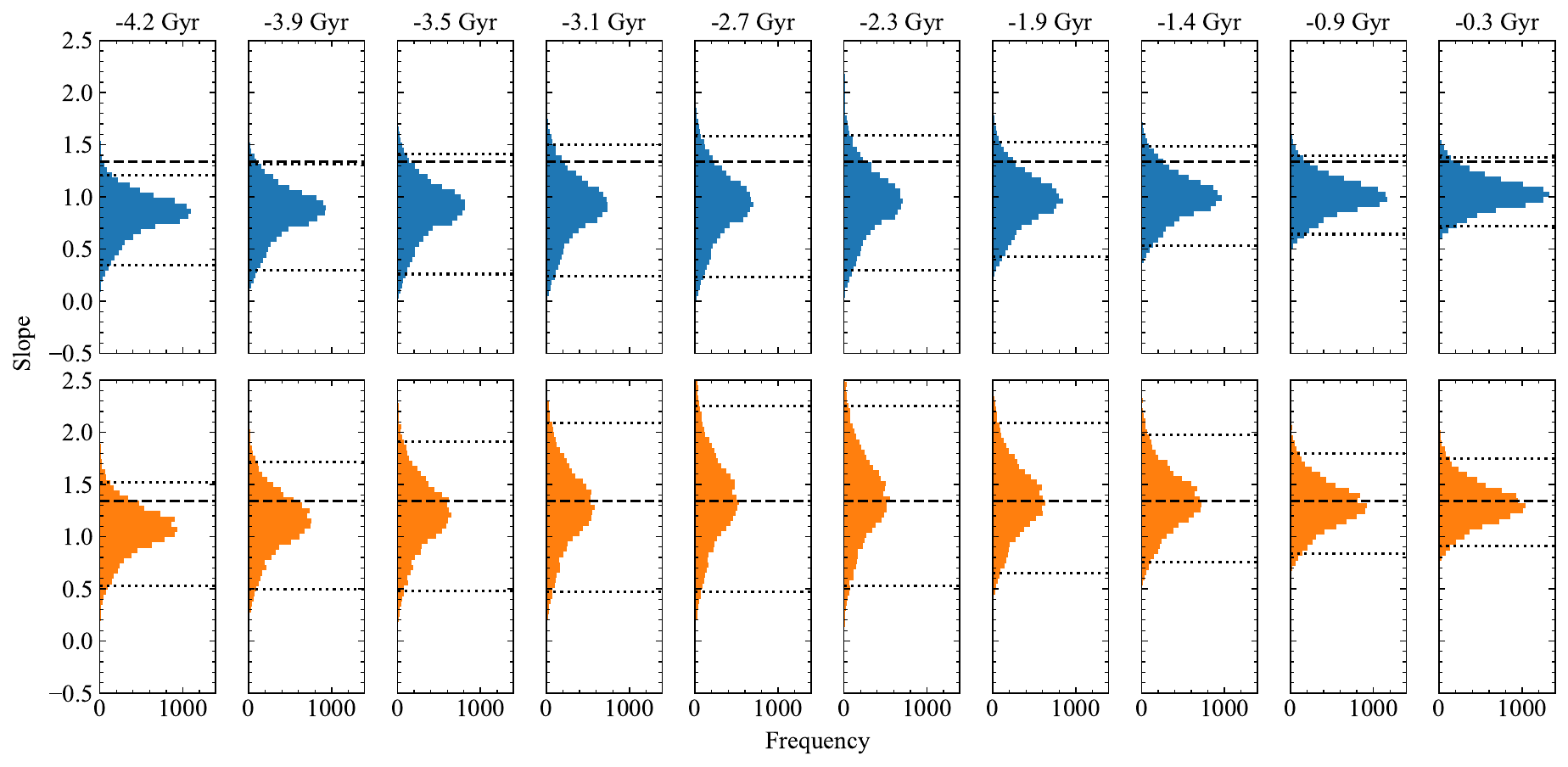}
    \caption{Slope distribution of the silicon isotope correlation line based on varying the nuclear reaction rates within the described uncertainties. Each histogram was calculated a given time in \ac{gce} history prior to solar system formation. Dotted lines represent the 95\% confidence intervals while the dashed line shows the measured slope by analyzing presolar SiC mainstream grains \citep{stephan2024pgd}. The top histograms (blue) show the slope distribution when using the \iso{28,29,30}Si($n,\gamma$) rates by \citet{guber}, the bottom histograms (orange) use the rates by \citet{bao}.}
    \label{fig:slope-variation-ocremove}
\end{figure}
Figure \ref{fig:slope-variation-ocremove} compares the variations of the slope of the silicon \ac{gce} line with the slope of the line predicted by mainstream SiC grains. The top row shows the slope predictions using the \iso{28,29,30}Si($n,\gamma$) rates by \citet{guber}, the bottom row uses the rates by \citet{bao}. Both sets of reaction rates were varied within their respective experimental uncertainties (Table~\ref{tab:rate-variation}). The slope at time $t$ is given by:
\begin{equation}
\frac{\del{29}{Si}{28,t+\delta t}-\del{29}{Si}{28,t}}{\del{30}{Si}{28,t+\delta t}-\del{30}{Si}{28,t}}.
\end{equation}
where $\delta t$ is the size of the time step of the \texttt{OMEGA+} simulation. The dashed line represents the slope of the presolar SiC grain mainstream correlation line of 1.342 \citep{stephan2024pgd}. The dotted lines show the 95\% confidence intervals of our simulations. Using the \iso{28,29,30}{Si}$(n,\gamma$) reaction rates by \citet{guber} (top part of Figure~\ref{fig:slope-variation-ocremove}), we find that the {slope of the} measured presolar SiC correlation line can be explained within the 95\% confidence interval for all simulations younger than 3.5\,Gyr prior to solar system formation. Moreover, the largest slope variation occurs at 2.7 Gyr prior to solar system formation. The majority of presolar SiC grains traveled through the interstellar medium {for short periods of time, e.g., up to 100\,Myr \citep{Heck2020}. Only large and very rare presolar SiC grains were measured in the study of \citet{Heck2020}. In contrast, our model data comparison is mostly with measurements of smaller, micrometer-sized grains, for which no measured ages exist. Calculated survival times in the interstellar medium are short \citep{jonesDustDestructionISM2011}, thus making an assumed $100$\,Myr lifetime for presolar grains a valid and conservative estimate.} In addition, the silicon isotopes represent the original composition that the grains' parent star started out with. Most mainstream grains likely originated from \ac{agb} stars with initial masses of $2-3\,M_\odot$ \citep[e.g.,][]{lugaroIsotopicCompositionsStrontium2003}. Assuming a simple mass-luminosity relation, we would thus expect the parent stars of these grains to have lived for {about} 0.6\,{Gyr} to 1.8\,{Gyr}. This roughly corresponds to \ac{gce} time steps between $-1.9\,$Gyr and $-0.9$\,Gyr in Figure~\ref{fig:slope-variation-ocremove}. {As shown in Figure~\ref{fig:slope-variation-ocremove}, the spread in the predicted slope is large enough in this time span of interest to include the observed slope within the 95\% confidence interval.} Therefore, our model shows that nuclear reaction rate uncertainties can indeed explain the discrepancies between measurements and models. This conclusion is even stronger when using the reaction rates by \citet{bao} (bottom part of Figure~\ref{fig:slope-variation-ocremove}). Here, all simulations agree within the 95\% confidence interval with the measured presolar grain slope. For the time range between -1.4\,Ga and -3.1\,Ga, even the median slope agrees excellently between models and data.

Our results show that nuclear uncertainties significantly impact the \ac{gce} of silicon isotopes. In particular, we can reproduce the slope of the mainstream line within nuclear uncertainties.  Our result however especially show the importance of the chosen \iso{28,29,30}{Si}$(n,\gamma)$ reaction rates. The \iso{30}{Si}$(n, \gamma)$\iso{31}{Si} rate given by \cite{guber} is 85\% lower than the rate provided by \citet{bao} at $kT = 8\ \text{keV}$, which is similar to the temperature that stars reach in our simulations. 
Similarly, the \iso{30}{Si}$(n, \gamma)$\iso{31}{Si} rate by \cite{beer:02} would be about a factor of two higher than \citet{guber}, somewhere in between the rate used in the stellar simulations considered here. Those variations are well above the experimental uncertainties that we used in the individual cases, i.e., 2$\sigma$ errors given by their reference sources, and our comparison between the rates of \citet{guber} and \citet{bao} clearly shows that the overall uncertainty in this rate is crucial to understand. In our scenario, the rate by \citet{bao} 
seems to be preferred with respect to the one by \citet{guber}, since it leads to an excellent agreement with the presolar SiC grain data. {\citet{zinner:06} already concluded that using the \citet{bao} nuclear cross sections results in a steeper slope in the silicon isotope \ac{gce} line than when using the rates by \citet{guber}. However, \citet{zinner:06} also concluded that the \citet{guber} nuclear reaction rates agree better with the \iso{30}{Si} excesses in SiC Z grains. These grains likely originated from \ac{agb} stars. The ratios of the neutron capture rates between \citet{guber} and \citet{bao} are 0.81 and 0.15 for \iso{29}{Si}$(n,\gamma)$ and \iso{30}{Si}$(n,\gamma)$ at 8\,keV, respectively. Simply from this ratios, it is clear that the rates by \citet{bao} will predict a steeper slope for the GCE correlation line and less \iso{30}{Si} excesses compared to the rates by \citet{guber}. While \citet{zinner:06} prefers the rates by \citet{guber} for the excess in \iso{30}{Si}, they also state that other factors such as the temperature at the bottom of the He intershell and the mass loss rate will significantly affect their conclusions. A future re-evaluation of the comparison by \citet{zinner:06} with newer AGB star models \citep[e.g.,][]{busso:2021} would be desirable.}

Our study suggests that a detailed evaluation of nucleosynthesis in O-C shell mergers is needed since it significantly affects the GCE of silicon isotopes. \citet{ritter-letter} 
proposed that the observed galactic trends of the odd-Z elements can be reproduced by increasing the O-C shell merger contributions in \ac{gce}. Our study shows that silicon isotopes in mainstream grains are also very %
sensitive to the nucleosynthesis in the O-C shell merger. %
Based on stellar models using the \citet{guber} silicon neutron-capture cross sections by \cite{Ritter_2018}, the occurrence of O-C shell mergers %
overproduce \iso{30}{Si}, which is in stark contrast to presolar SiC mainstream grain measurements. {On the other hand and in contrast to all other \ac{gce} models we looked at, the O-C shell merger models predict the correct solar \iso{29}Si/\iso{28}{Si}}. The contradicting evidence highlights the importance of a further investigation of both the relevant nuclear reaction rates discussed here and the dynamics and nucleosynthesis of the O-C shell mergers and their effects on \ac{gce}. Isotopic studies %
represent a finer signature to study these effects than elemental abundances.

\section{Conclusions} \label{sec:conc}

We presented %
{an impact study of} the effects of nuclear reaction rate uncertainties on the \ac{gce} of silicon isotopes and compared our results with measurements of presolar SiC mainstream grains. While currently no \ac{gce} model can adequately describe {the measured isotope ratios and their distribution}, we found for the first time that the slope of the simulated silicon isotope correlation agrees well with measurements of presolar SiC mainstream grains within nuclear reaction rate uncertainties. The slope of the silicon \ac{gce} trend agrees with the measurements within the 95\% confidence interval when choosing the \iso{28,29,30}Si$(n,\gamma)$ rates by \citet{guber}. Using the higher rates by \citet{bao} shows even better agreement, in fact, the expected average slope using these reaction rates agrees perfectly with the data. 

Based on the stellar models available for this study, our results show that massive stars undergoing an O-C shell merger %
{allow to reproduce the solar $^{29}$Si ratio with GCE simulations, but overproducing $^{30}$Si compared to the observations.} 
{ \citet{ritter-letter} found that} a significant contribution {from O-C mergers would be a possible solution} %
to solve the underproduction of odd-Z elements in the Galaxy. {In this work we show that in the case of Si isotopes the impact of O-C shell mergers is also relevant, but it is controversial}. %

Using \citet{bao} or \cite{beer:02} rates {for the Si neutron-capture cross sections}, \ac{ccsn} models (and GCE models) would clearly predict lower amounts of \iso{30}Si{, which could potentially mitigate the discrepancy between %
GCE simulations and observations.}
{Our findings thus underline the importance of future precision reaction rate measurements in order to finally solve this conundrum. }
New experimental data for these rates are paramount to define the role of O-C shell mergers in GCE.
Isotope measurements %
are a very sensitive probe to study various stellar processes, and should be considered %
in future \ac{gce} studies of the solar neighborhood. %

\vspace*{1cm}
We would like to thank James Cho for helpful discussion.
MP thank the support from the NKFI via K-project 138031 and the Lend\"ulet Program LP2023-10 of the Hungarian Academy of Sciences. We acknowledge the support to NuGrid from JINA-CEE (NSF Grant PHY-1430152), and ongoing access to {\tt viper}, the University of Hull High Performance Computing Facility. MP and RT acknowledges the support from the European Union’s Horizon 2020 research and innovation programme (ChETEC-INFRA -- Project no. 101008324), and the IReNA network supported by US NSF AccelNet (Grant No. OISE-1927130).
\vspace*{1cm}

\software{\texttt{matplotlib} \citep{Hunter:2007}, \texttt{NumPy} \citep{2020NumPy}, \texttt{SciPy} \citep{2020SciPy-NMeth}, \texttt{pandas}} \citep{reback2020pandas,mckinney-proc-scipy-2010}

\clearpage
\appendix

\section{Mass coordinates of relevant silicon production zones}\label{app:o-burning}
Table~\ref{tab:explosive-zone} shows the mass coordinates of the O-burning zone in all massive star models by \citet{Ritter_2018}. 
\begin{table}[hbt]
    \centering
    \caption{Mass coordinates of stellar zones where explosive oxygen burning takes place in stellar models by \citet{Ritter_2018} of given mass and metallicity.}
\begin{tabular}{ccc}
        \hline
          Mass ($\msun$)    & Metallicity  &  Mass Coordinate Range ($\msun$)\\
        \hline
        12 & 0.01 & 2.12 - 2.78\\
        12 & 0.02 & 1.96 - 2.43\\
        25 & 0.01 & 6.05 - 9.19\\
        25 & 0.02 & 6.55 - 8.96\\
        \hline
    \end{tabular}
    \label{tab:explosive-zone}
\end{table}

\section{Trajectories}\label{app:trajectories}
\begin{figure}[tb]
    \centering
\includegraphics[width=0.6\textwidth]{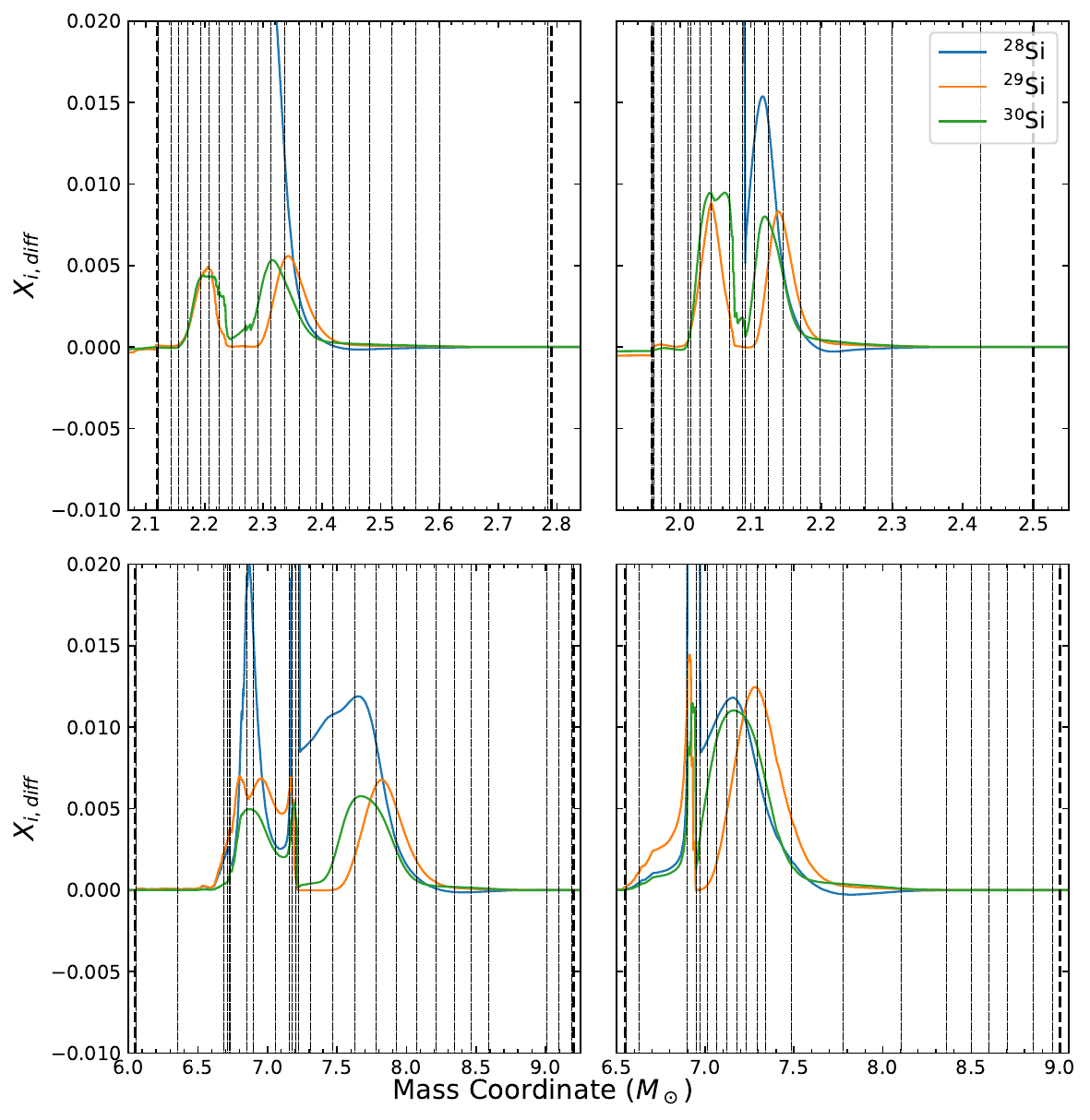}
    \caption{{Isotopic abundances of silicon isotopes at given mass coordinate for} representative 1D trajectories taken for the $12\,M_\odot$ (top) and $25\,M_\odot$ (bottom) stars for $Z=0.01$ (left) and $Z=0.02$ (right).}
    \label{app:grid-point}
\end{figure}

For all stellar models, we take representative 1D temperature, plasma-density trajectories from the respective silicon production zones, ranging from mass-coordinates $m_a$ to $m_b$. The subscript represents the index of the computational grid of the \texttt{MESA} stellar evolution code, where 0 represents the center of the star. For all stellar models, 21 trajectories equally spaced in index space are taken. For stellar model with $M/\msun = 25,\ Z = 0.01$, 5 additional trajectories equally spaced in index space are taken between mass-coordinate 6.852 and 7.219 $\msun$ to capture all the silicon isotopes production peaks. The dashed lines in Figure \ref{app:grid-point} shows the representative trajectories taken. The solid lines represent the difference between the post-supernova and pre-supernova silicon isotopic abundances. 

\section{Convergence of MC simulations}\label{app:mc_convergence}

\begin{figure}
    \centering
    \includegraphics[width=0.6\textwidth]{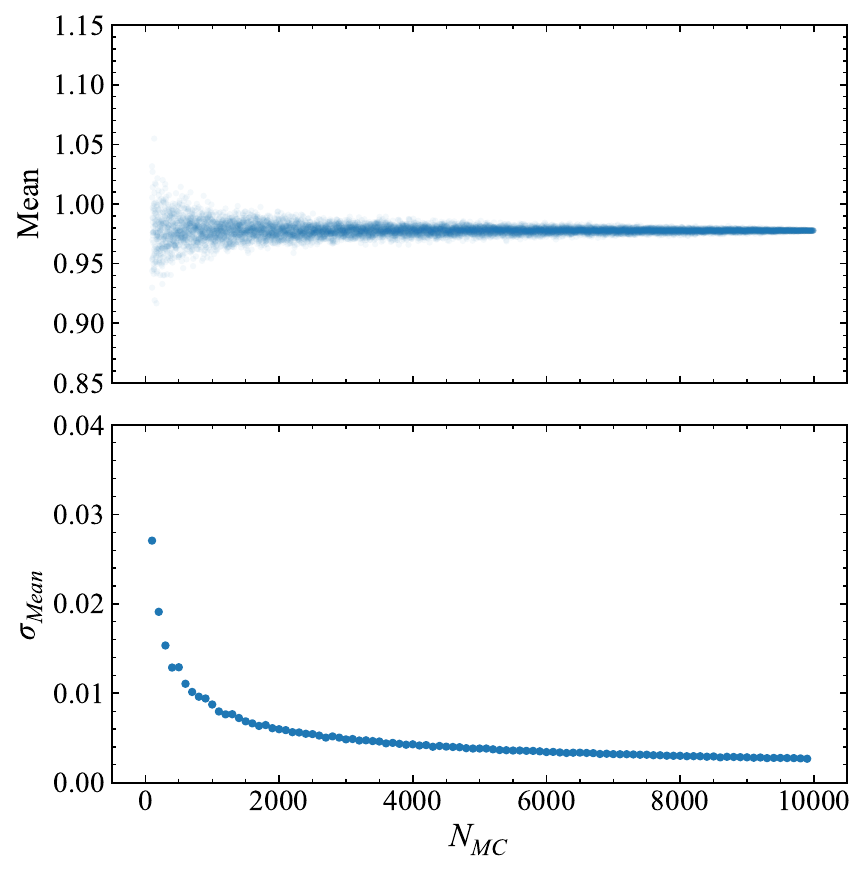}
    \caption{Convergence test of our \ac{mc} model. Clearly, the mean silicon isotope correlation lines has converged well at our chosen value after several thousand \ac{mc} runs, showing that our results will not change when running more than 10,000 simulations.}
    \label{fig:convergence}
\end{figure}

The convergence of the \ac{mc} simulations is tested using the distribution of the silicon isotope correlation lines at $t=-2.3\,$Gyr with increasing sample size $N_{MC}$. Figure \ref{fig:convergence} shows the means of the distribution with sample size $N_{MC}$, where $N_{MC}\in \{100,110,120,...,10000\}$. We see that the estimated mean stabilize as $N_{MC}$ increases. We estimate the point of convergence through a bootstrapping technique, a statistical procedure that generates a set of simulated samples by sampling from an existing sample with replacement. It is generally used to estimate the accuracy of a sample estimates such as mean and median. The exact procedure we take is as following. First, we take a sample of size $N_{MC}\in \{100,200,300,...,10000\}$ from the 10000 \ac{mc} results. Then, we generate 10000 simulated samples of size $N_{MC}$ by sampling from the sample of \ac{mc} results with replacement and calculate the mean of each simulated sample. Finally, we calculate the standard deviation of the means of the 10000 simulated samples. Figure \ref{fig:convergence} shows the calculated standard deviations for all $N_{MC}$. The estimated standard deviations of the sample means monotonically decrease as $N_{MC}$ increase, suggesting the \ac{mc} simulations we performed is converging. When $N_{MC} = 10000$, the standard deviations of the sample mean estimated by the bootstrapping method is $\sim 0.003$ which is $<0.5\%$ of the sample mean suggesting the convergence of the \ac{mc} simulation at $N_{MC} = 10000$.

\section{Silicon production variations within the relevant zones}\label{app:silicon_production_variation}

\begin{figure}[tb]
    \centering
    \includegraphics[width=0.6\textwidth]{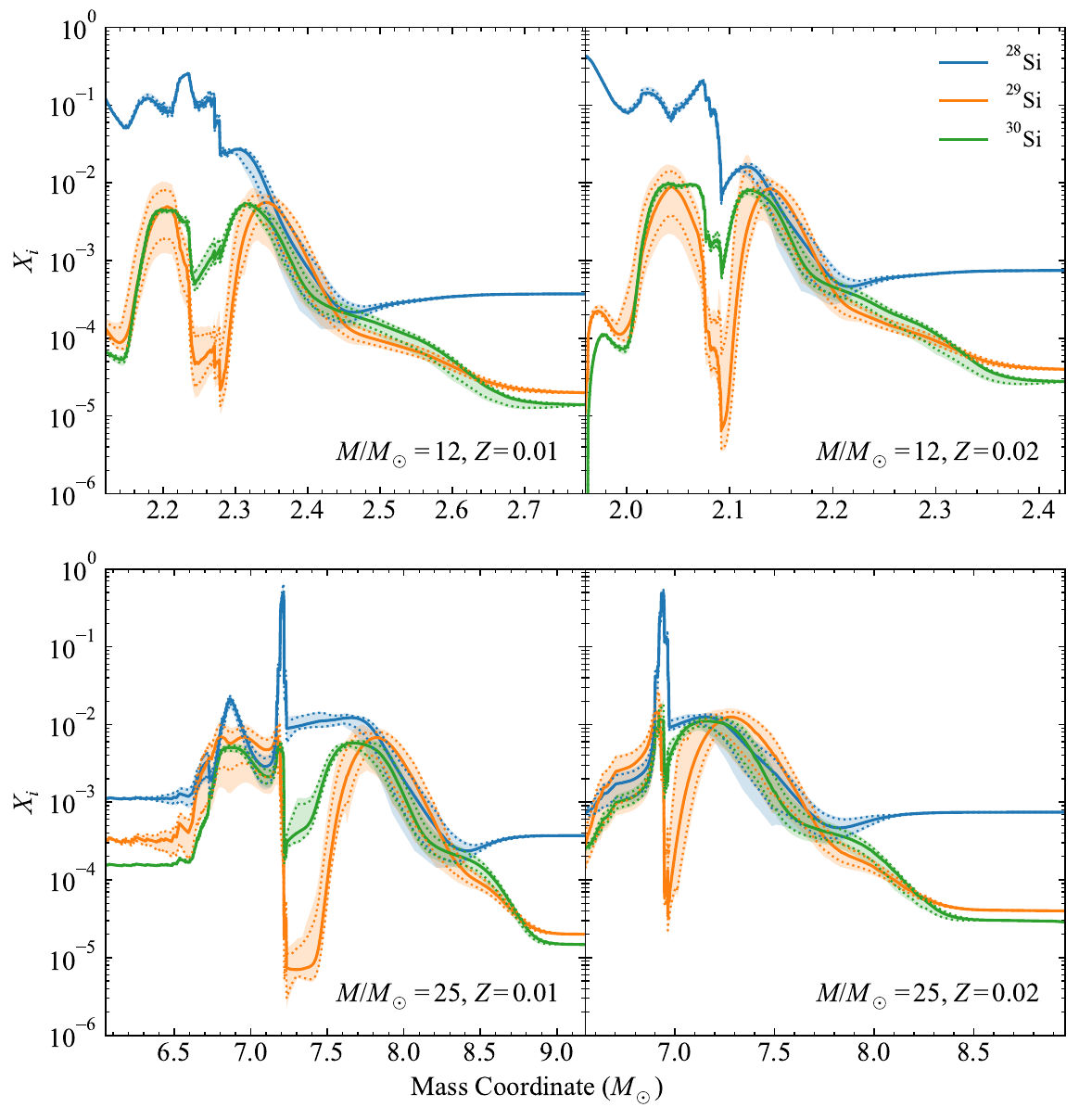}
    \caption{Silicon abundance at the end of the \ac{ccsn} explosion within the explosive silicon production zone of the $12\,M_\odot$ (top) and $25\,M_\odot$ (bottom) stars for $Z=0.01$ (left) and $Z=0.02$ (right). Lines represent the unmodified abundance, as in \citet{Ritter_2018}. Shaded areas show the overall range for 10000 \ac{mc} simulations while varying the nuclear reactions rates, while the dotted lines show the 95\% confidence interval of our \ac{mc} simulations.}
    \label{fig:stellar-variation}
\end{figure}

Figure \ref{fig:stellar-variation} shows the variations of the post-supernova silicon isotopic abundances in the silicon production zones for selected massive star models. The solid lines represent the post-supernova compositions in the default stellar nucleosynthesis models \citep{Ritter_2018}, i.e., without reaction rate variations. The shaded areas show the whole range of variance as determined by our \ac{mc} simulations, with dotted contours marking the 95\% confidence intervals. The largest variation shows up in the results for \iso{29}{Si} due to large uncertainties in the nuclear reaction rates. Final isotopic abundances in the oxygen burning zones can vary more than an order of magnitude. On the other hand, \iso{28}{Si} and \iso{30}{Si} show less variation due to reaction rate uncertainties.

\clearpage

\bibliographystyle{aasjournal}
\bibliography{references}

\end{document}